\documentclass[sn-standardnature]{sn-jnl}

\jyear{2023}%

\theoremstyle{thmstyleone}%
\theoremstyle{thmstyletwo}%

\theoremstyle{thmstylethree}%

\begin{document}

\title[Extrasolar radiation belt]{Resolved imaging of an extrasolar radiation belt around an ultracool dwarf}


\author*[1,2,3,4]{\fnm{Melodie M.} \sur{Kao}}\email{mmkao@ucsc.edu}
\equalcont{These authors contributed equally to this work.}

\author[5]{\fnm{Amy J.} \sur{Mioduszewski}}\email{amiodusz@nrao.edu}
\equalcont{These authors contributed equally to this work.}

\author[6]{\fnm{Jackie} \sur{Villadsen}}\email{jrv012@bucknell.edu}

\author[2]{\fnm{Evgenya L.} \sur{Shkolnik}}\email{shkolnik@asu.edu}

\affil*[1]{\orgdiv{Department of Astronomy \& Astrophysics}, \orgname{University of California, Santa Cruz}, \orgaddress{\street{1156 High St}, \city{Santa Cruz}, \postcode{95064}, \state{CA}, \country{USA}}}

\affil[2]{\orgdiv{School of Earth \& Space Exploration}, \orgname{Arizona State University}, \orgaddress{\street{781 Terrace Mall}, \city{Tempe}, \postcode{85287}, \state{AZ}, \country{USA}}}

\affil[3]{\orgdiv{51 Pegasi b Fellow}, \orgname{Heising-Simons Foundation}}
\affil[4]{\orgdiv{Hubble Prize Postdoctoral Fellow}, \orgname{NASA}}

\affil[5]{ \orgname{National Radio Astronomy Observatory}, \orgaddress{\street{1003 Lopezville Rd}, \city{Socorro}, \postcode{87801}, \state{NM}, \country{USA}}}

\affil[6]{ \orgdiv{Department of Physics \& Astronomy}, \orgname{Bucknell University}, \orgaddress{\street{1 Dent Drive}, \city{Lewisburg}, \postcode{17837}, \state{PA}, \country{USA}}}



\maketitle

\clearpage
\textbf{Radiation belts are present in all large-scale Solar System planetary magnetospheres: Earth, Jupiter, Saturn, Uranus, and Neptune \citep{maukFox2010}. These persistent equatorial zones of trapped high energy particles up to tens of MeV  \citep{maukFox2010} can produce bright radio emission \citep{Bolton2002Natur.415..987B, Bolton2004, Kollmann2018JGRA..123.9110K} and impact the surface chemistry of close-in moons \citep{Gudipati2021NatAs...5..276G}. Recent observations confirm planet-like radio emission such as aurorae from large-scale magnetospheric current systems \citep{Nichols2012ApJ...760...59N, Turnpenney2017MNRAS.470.4274T, Saur2021AA...655A..75S} on very low mass stars and brown dwarfs \citep{Hallinan2015Natur.523..568H, Kao2016ApJ...818...24K, Pineda2017ApJ...846...75P}. These objects, collectively known as ultracool dwarfs, also exhibit quiescent radio emission hypothesized to trace stellar coronal flare activity \citep{Williams2014ApJ...785....9W} or extrasolar radiation belt analogs \citep{Hallinan2006ApJ...653..690H, Pineda2017ApJ...846...75P, Kao2019MNRAS.487.1994K, Leto2021MNRAS.507.1979L, Climent2022AA...660A..65C}. Here we present high resolution imaging of the ultracool dwarf LSR~J1835+3259 demonstrating that this radio emission is spatially resolved  and traces a long-lived, double-lobed, and axisymmetric structure similar in morphology to the Jovian radiation belts. Up to 18 ultracool dwarf radii separate the two lobes. This structure is stably present in three observations spanning $>$1 year. We infer a belt-like distribution of plasma confined by the magnetic dipole of LSR~J1835+3259, and we estimate $\sim$15 MeV electron energies that are consistent with those measured in the Jovian radiation belts \citep{Kollmann2018JGRA..123.9110K}. Though more precise constraints require higher frequency observations, a unified picture where radio emissions in ultracool dwarfs manifest from planet-like magnetospheric phenomena has emerged. }

Historically, stars with strong magnetic heating in their atmospheres have informed interpretations of ultracool dwarf quiescent radio emissions \citep{Williams2014ApJ...785....9W}. In particular, a tight correlation between the non-thermal quiescent radio and thermal X-ray luminosities from magnetically active stars \citep{Gudel1993ApJ...405L..63G} also holds for solar and stellar flare emission \citep{Benz1994A&A...285..621B}, suggesting that flare-accelerated electrons produce quasi-steady quiescent radio emission from active stars. However, ultracool dwarf quiescent radio luminosities do not follow empirical flare relationships \citep{Williams2014ApJ...785....9W, Pineda2017ApJ...846...75P}. Instead, a mechanism entirely different from low-level flaring may drive the quiescent emission. 

As an alternative, Jovian radio emissions have inspired recent studies \citep{Pineda2017ApJ...846...75P, Kao2019MNRAS.487.1994K, Leto2021MNRAS.507.1979L, Climent2022AA...660A..65C}. The three main sources of Jovian radio emission are thermal cloud emission from the photosphere \citep{dePater1982, dePater2016}, aurorae manifested as circularly polarized and rotationally periodic bursts  powered by the electron cyclotron maser instability \citep{BurkeFranklin1955JGR....60..213B, Zarka1998JGR...10320159Z}, and more gradually varying quiescent synchrotron emission from high-energy electrons populating radiation belts extending up to 13 Jupiter radii (R$_{\mathrm{J}}$) in the Jovian dipole field \citep{Bolton2004}. 

For ultracool dwarfs, gigahertz (GHz) flux densities at hundreds of microjanskies ($\mu$Jy) rule out thermal emission, which should be sub-$\mu$Jy from their $\lesssim$2800~K  photospheres.  Instead, their slowly varying quiescent radio emissions point to persistent populations of relativistic electrons in their magnetospheres \citep{Osten2009ApJ...700.1750O, Williams2013ApJ...767L..30W, Kao2016ApJ...818...24K, kao2023a}.  Intriguingly, the quiescent radio luminosities of auroral ultracool dwarfs correlate well with their Balmer emission  \citep{Pineda2017ApJ...846...75P, Richey-Yowell2020} tracing auroral rather than the usual chromospheric activity \citep{Hallinan2015Natur.523..568H, Kao2016ApJ...818...24K}. This suggests that conditions enabling auroral processes in ultracool dwarfs \citep{Nichols2012ApJ...760...59N, Turnpenney2017MNRAS.470.4274T, Saur2021AA...655A..75S} may also support strong quiescent radio emission. 

A search for companions around an auroral ultracool dwarf marginally resolved 8.5~GHz emission \citep{Forbrich2009ApJ...706L.205F}, which we propose  hints at a possible radiation belt like Jupiter's. However, at 10.6 pc distant, current instrumentation cannot conclusively resolve the spatial distribution of its emitting electrons.

\begin{table}[t]
\begin{center}
\begin{minipage}{0.8\textwidth}
\caption{Properties of LSR~J1835+3259}\label{tab:target}
\begin{tabular*}{\textwidth}{@{\extracolsep{\fill}}llcll@{\extracolsep{\fill}}}
\toprule
    Property        &
            &
            &
    units   &
 	ref       \\
\midrule	
    RA                                  &                            & 18 35 37.8801226673                     & J2000             & \citep{Gaia2020yCat.1350....0G} \\
    Dec                                 &                            & +32 59 53.316099140                     & J2000             & \citep{Gaia2020yCat.1350....0G} \\
    $\pi$                               &                            &  175.7930 $\pm$ 0.0468                  & mas               & \citep{Gaia2020yCat.1350....0G} \\
    d                                   &                            & 5.6875 $\pm$ 0.002927                   & pc                & \citep{Gaia2020yCat.1350....0G} \\
    $\mu_{\alpha}\cos(\delta)$          &                            & $-$72.650  $\pm$ 0.047                  & mas yr$^{-1}$     & \citep{Gaia2020yCat.1350....0G} \\
    $\mu_{\delta}$                      &                            & $-$755.146 $\pm$ 0.052                  & mas yr$^{-1}$     & \citep{Gaia2020yCat.1350....0G} \\
    SpT                                 &                            & M8.5V                                   &                   & \citep{Deshpande2012AJ....144...99D} \\
    $R$                                 &                            & 1.07 $\pm$ 0.05                         & R$_{\mathrm{J}}$  & \citep{Filipazzo2015ApJ...810..158F} \\
    $\tau$                              &                            &  $\geq$500                              & Myr               & \citep{Filipazzo2015ApJ...810..158F}  \\
    $M$                                 &                            &  77.28 $\pm$ 10.34                      & M$_{\mathrm{J}}$  & \citep{Filipazzo2015ApJ...810..158F}\\
    $P_{\mathrm{rot}}$                  &                            & 2.845 $\pm$ 0.003                       & hr                & \citep{Harding2013ApJ...779..101H} \\
    $i$                                 &                            &  90                                     & $^{\circ}$        & \citep{Filipazzo2015ApJ...810..158F, Harding2013ApJ...779..101H}  \\
    $F_{\nu}$\footnotemark[1]           &  8.46 GHz                  & 464 $\pm$ 10                            & $\mu$Jy           & \citep{Berger2008ApJ...676.1307B} \\
                                        &  8.44 GHz                  & 722 $\pm$ 15                            & $\mu$Jy           & \citep{Hallinan2008ApJ...684..644H} \\
                                        &   ---\,\,\,\,\, GHz                 & 525 $\pm$ 15                            & $\mu$Jy           & \citep{Berger2006ApJ...648..629B} \\
                                        &  97.5 GHz                  & 114 $\pm$ 18                            & $\mu$Jy           & \citep{hughes2021AJ....162...43H} \\
\botrule
\end{tabular*}
\footnotetext{LSR~J1835+3259 is not a known binary.}
\footnotetext[1]{Stokes I quiescent radio flux density reported in the literature. When available, listed observing frequencies are the center of the reported observing band.}
\end{minipage}
\end{center}
\end{table}

Instead, the nearby ($5.6875\pm0.00292$ pc \citep{Gaia2020yCat.1350....0G}) ultracool dwarf LSR~J1835+3259 is an ideal target. Its periodically bursting ($P=2.84 \pm 0.01$ hr) \citep{Hallinan2008ApJ...684..644H} 8.4 GHz radio aurorae trace $\geq$3~kiloGauss  magnetic fields near its surface \citep{Hallinan2015Natur.523..568H}, and magnetically sensitive 819 nm sodium emission lines indicate 5.1~kiloGauss magnetic fields averaged over 11\% of its surface \citep{Berdyugina2017ApJ...847...61B}.  LSR~J1835+3259 also emits quiescent radio emission at the same 8.4~GHz frequencies \citep{Berger2006ApJ...648..629B, Hallinan2008ApJ...684..644H, Berger2008ApJ...676.1307B} that slowly varies in average flux density by $\sim$50\%, as well as faint 97.5 GHz emission \citep{hughes2021AJ....162...43H}. With no detectable infrared excess indicating a disk \citep{Avenhaus2012AA...548A.105A} and periodic Balmer line emission associated with aurorae \citep{Hallinan2015Natur.523..568H} rather than accretion, its 97.5 GHz emission may originate from the same population of relativistic electrons as its 8.4 GHz quiescent emission.

\begin{figure}[t]
\centering
\includegraphics[width=1\textwidth]{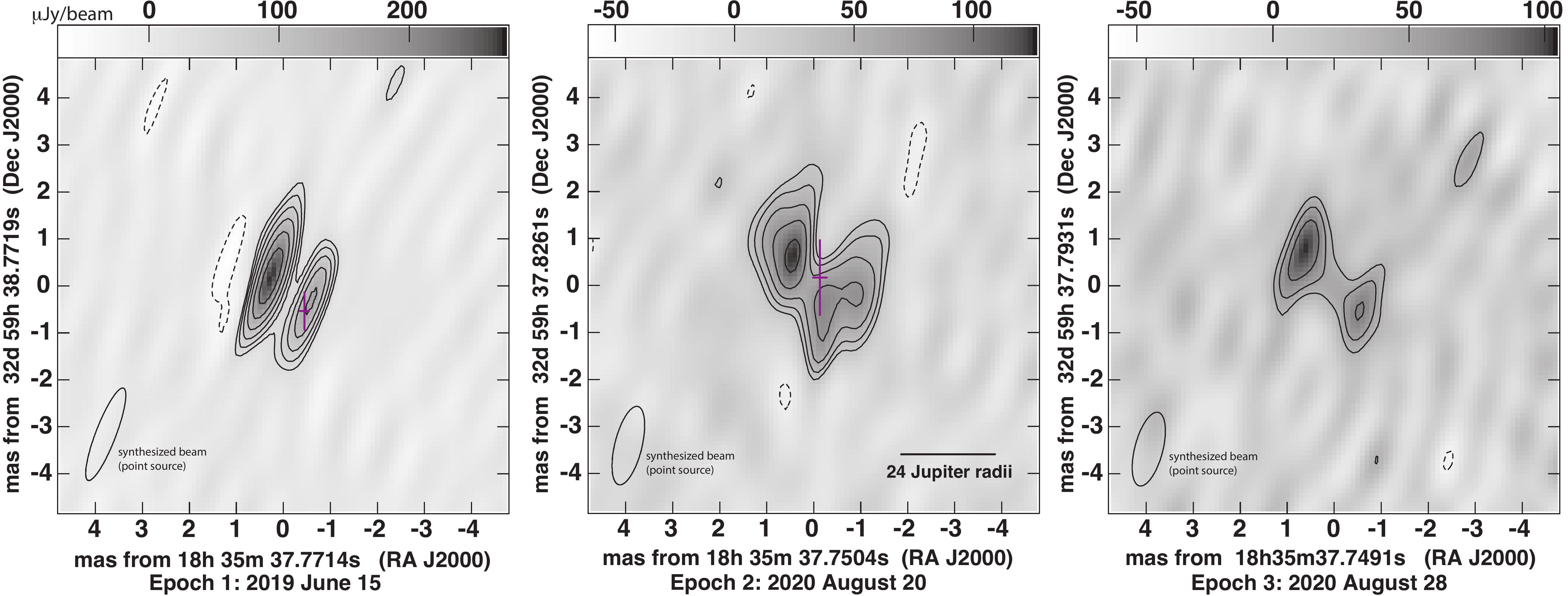} 
\caption{Extended emission at 8.4 GHz from LSR~J1835+3259 with auroral emission excised and contours in 3$\sigma_{\mathrm{rms}} \, \times$ (-1, 1, $\sqrt{2}$, 2, 2$\sqrt{2}$, 4)  increments.  The synthesized beam (black ellipse) sets the resolution element for each image and appears shortened along one axis due to the array configuration. Crosshairs indicate aurorae centroids and 3$\sigma$ positional errors (magenta). Coordinates are for midnight in International Atomic Time. }
\label{fig:images}
\end{figure}

LSR~J1835+3259 straddles the substellar boundary. Its M8.5 spectral type \citep{Deshpande2012AJ....144...99D} corresponds to a $2316 \pm 51$~K effective temperature \citep{Filipazzo2015ApJ...810..158F}. Modeling finds that for an age $\geq$500~million years (Myr), it will have a mass of $77.28 \pm 10.34$~Jupiter masses ($M_{\mathrm{J}}$) near the hydrogen burning limit that differentiates between low-mass stars and massive brown dwarfs  \citep{Filipazzo2015ApJ...810..158F}. These correspond to a radius $R_{\mathrm{UCD}} = 1.07 \pm 0.05$~Jupiter radii ($R_\mathrm{J}$) that, together with its $2.845\pm0.003$~hr optical rotation period and projected surface velocity $v \sin i = 50 \pm 5$~km~s$^{-1}$ \citep{Harding2013ApJ...779..101H}, imply that LSR~J1835+3259 is edge-on relative to our line of sight with a rotation axis inclined at an angle $i \approx 90^{\circ}$. 

Using the  High Sensitivity Array (HSA) of 39 radio dishes spanning the USA to Germany, we searched for and imaged extended quiescent radio emission at 8.4 GHz from  LSR~J1835+3259 from a large-scale magnetospheric plasma structure as evidence of an extrasolar analog of Jovian radiation belts. Our observing campaign consisted of three 5-hour epochs from 2019 to 2020 (Table~\ref{tab:spatial}).

We find that quiescent radio emissions from LSR~J1835+3259 persist throughout each epoch and exhibit a double-lobed morphology that is stable for more than one year (Figure \ref{fig:images}). Up to $18.47 \pm 2.20$~$R_{\mathrm{UCD}}$ separate its radio lobes, which have no detectable circular polarization  (Tables \ref{tab:imaging} and \ref{tab:spatial}). These data constitute the first resolved radio imaging of an ultracool dwarf magnetosphere.

\begin{figure}[t]
\centering
\includegraphics[width=0.45\textwidth]{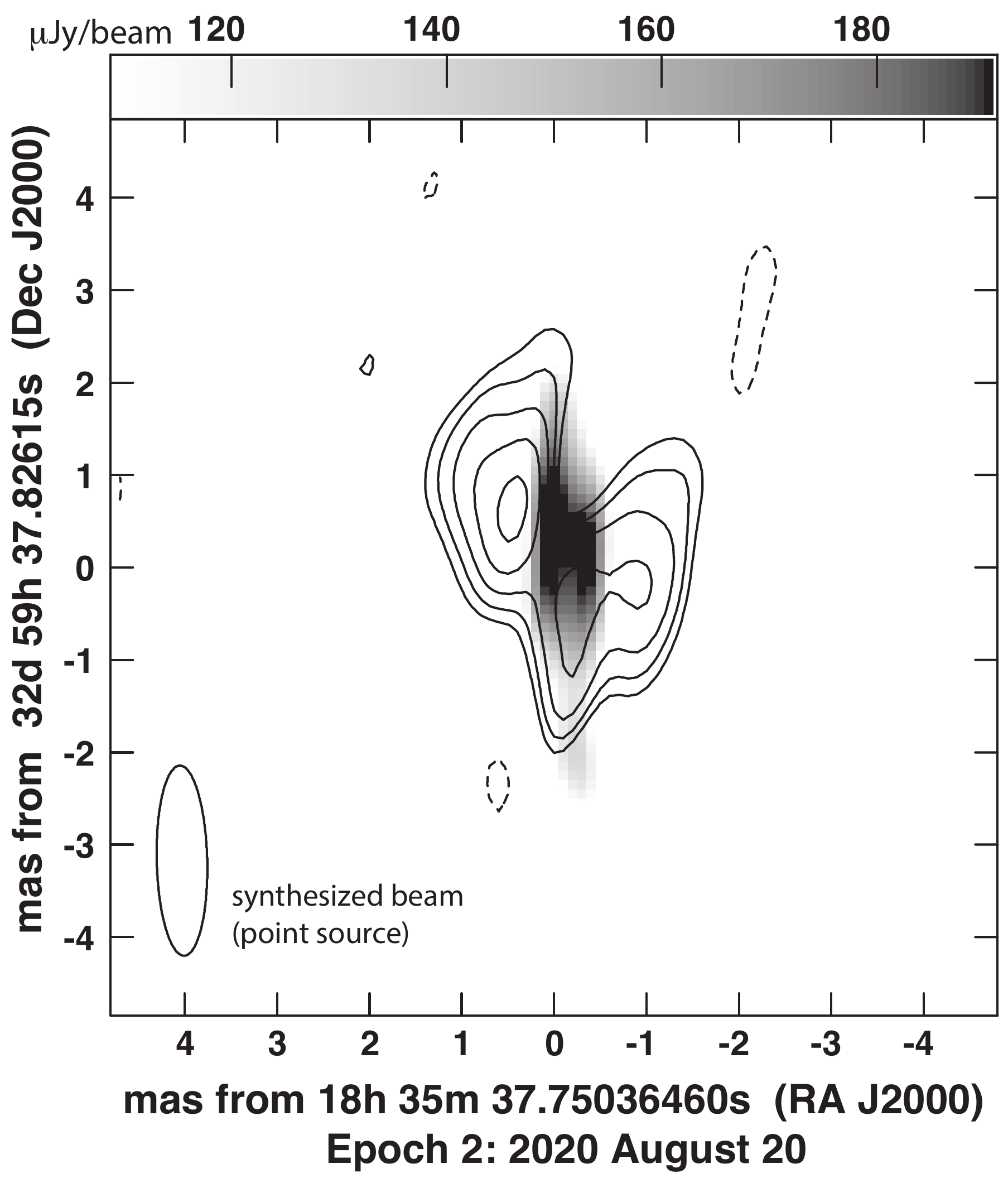} 
\caption{ 3$\sigma_{\mathrm{rms}} \, \times$ (-1, 1, $\sqrt{2}$, 2, 2$\sqrt{2}$, 4) contours of quiescent 8.4 GHz emission from LSR\,J1835+3259 in Epoch 2 with its right circularly polarized aurora overlaid in grey scale.  The synthesized beam (black ellipse) sets the resolution element for the aurora and is determined by the array configuration. Figure \ref{fig:images} shows synthesized beams for quiescent emission. The aurora appears centrally located with respect to the double-lobed morphology of the quiescent emission. Coordinates are for midnight in International Atomic Time. } 
\label{fig:aurora+belt}
\end{figure}

Auroral bursts appear centrally located between the two lobes in Epoch 2, which has the highest-quality data. Figure \ref{fig:aurora+belt} shows an aurora from Epoch~2 separately imaged and then overlaid on the quiescent emission contours  from that same epoch. Epoch~1 cannot detect the faint extended emission observed in later epochs due to missing antennas, so aurorae appear coincident with the right quiescent radio lobe (Figure \ref{fig:images}).  We simulate an observation of the Epoch~2 image using the Epoch~1 antenna configuration and find that the lobe separation in Epoch 1 is consistent this simulated Epoch 2 observation. The relative locations between aurorae and quiescent radio lobes are also consistent. In Epoch~3, aurorae are too faint to confidently locate.

8.4~GHz aurorae originate in 3~kiloGauss magnetic fields near the surface of LSR~J1835+3259 \citep{Hallinan2008ApJ...684..644H}. From Epoch 2 (Figure \ref{fig:aurora+belt}), we infer that lobe centroids sit at $6-9~R_\mathrm{UCD}$ from the ultracool dwarf (with 7-10\% uncertainties; Table~\ref{tab:imaging}), while their outer extents reach at least $16-18$~$R_\mathrm{UCD}$.  The structure may be even larger; individual epochs may not be sensitive to fainter and more extended emission, as is the case for Epoch 1. 

At these large extents, dipole magnetic fields decaying with radius as $B\propto r^{-3}$ will dominate higher-order magnetic fields ($B \propto r^{-4}$ or more rapid decay) of similar surface field strengths inferred for  LSR~J1835+3259 with multiwavelength spectra \citep{Berdyugina2017ApJ...847...61B, Kuzmychov2017ApJ...847...60K}. Indeed, the persistent double-lobed and axisymmetric morphology observed is consistent with a stable dipole magnetic field, and theoretical treatments assuming such can explain radio aurorae observed from LSR~J1835+3259 and other ultracool dwarfs \citep{Nichols2012ApJ...760...59N, Turnpenney2017MNRAS.470.4274T, Saur2021AA...655A..75S}. 

Our observations present compelling evidence for the first known analog of Jovian radiation belts outside of our Solar System, consisting of a long-lived population of relativistic electrons confined in a global magnetic dipole field \citep{RoussosKollmann2021GMS...259..499R}.  To explore implications of the lobe separation for electron energies, we consider a $\geq$3~kiloGauss surface dipole field. At the lobe centroids, the field strength and corresponding non-relativistic electron cyclotron frequency $\nu_{\mathrm{c}} = eB/2\pi m_ec$  would be 2~Gauss and 6 MHz.

First, the spatial extent of the quiescent radio emission is consistent with synchrotron emission from relativistic electrons.  An electron gyrating about a magnetic field emits at multiple harmonics $s$ of its relativistic cyclotron frequency, $\nu_c/\gamma$:
\begin{equation}
    \nu = s \left( \frac{\nu_c}{\gamma} \right) \,,
\end{equation}
where $\gamma$ is the Lorentz factor of the electron  described by its speed, and $\nu$ is the observed frequency \citep{Gudel2002ARAA..40..217G}.  Since $\gamma>1$,  8.5~GHz emission corresponds to $s\geq1500$ in the lobe centroids. Higher order magnetic fields fall off more rapidly in strength than dipole fields, resulting in higher harmonics. These high harmonics rule out gyrosynchrotron emission from mildly relativistic electrons, which typically emit at harmonics $s \approx 10-100$ \citep{Gudel2002ARAA..40..217G}. 

\begin{table}[t]
\begin{center}
\begin{minipage}{\textwidth}
\caption{Imaging: Quiescent spatial extents}\label{tab:spatial}
\begin{tabular*}{\textwidth}{@{\extracolsep{\fill}}lccccc@{\extracolsep{\fill}}}
\toprule%
Epoch: Date\footnotemark[1]                                 & 
Synth. Beam & \multicolumn{2}{@{}l@{}}{Centroid Sep. \& PA} & 
Left diameter\footnotemark[2]                               & 
Right diameter\footnotemark[2]                              \\
                                            &      
(mas $\times$ mas)                          & 
(mas)                                       &   
($^{\circ}$)                                & 
(mas)                                       & 
(mas)                                       \\
                        &                         
                        & 
($R_\mathrm{UCD}$)      &      
                        & 
($R_\mathrm{UCD}$)      & 
 ($R_\mathrm{UCD}$)     \\
\midrule
1: 2019 Jun 15      & 2.10$\times$0.43  & \,\,\,1.04$\pm$0.05      &  49.6             &  --               & -- \\
                    &                   & 11.52$\pm$0.81           &                   &  --               & -- \\
2: 2020 Aug 20      & 1.71$\times$0.58  & \,\,\,1.61$\pm$0.1       & 47.4              &  0.58$\pm$0.10    &  0.71$\pm$0.12  \\
                    &                   & 17.95$\pm$1.39           &                   &   6.44$\pm$1.19     &  7.87$\pm$1.38\\
3: 2020 Aug 28      & 1.61$\times$0.60  & \,\,\,1.66$\pm$0.15      &  50.0             &   0.66$\pm$0.11   & 0.83$\pm$0.19 \\
                    &                   & 18.47$\pm$1.85           &                   &  7.39$\pm$1.31      & 9.28$\pm$2.20            \\
\botrule
\end{tabular*}
\footnotetext{We use $R_\mathrm{UCD} = 1.07 \pm 0.05$ $R_\mathrm{J}$ \citep{Filipazzo2015ApJ...810..158F} and propagate radius uncertainties in reported dimensions.}
\footnotetext[1]{Observations combine the Very Long Baseline Array (VLBA), Karl G. Jansky Very Large Array (VLA), Greenbank Telescope (GBT), and Effelsberg Telescope (EB). For missing dishes, hyphenated pairs denote baselines and numbers denote hours missing when $<$5: \\ Epoch~1 - VLBA:2, VLBA Mauna Kea (MK), VLBA Saint Croix (SC), GBT;  \\ Epoch~2 - MK-EB;  Epoch~3 - MK.}  
\footnotetext[2]{Lobe diameters along the minor axes, which are resolved except for in Epoch 1 due to missing antennas.  Major axes are unresolved.}
\end{minipage}
\end{center}
\end{table}

Instead, such high harmonics indicate synchrotron emission from very relativistic electrons, which cannot produce strong circular polarization.  This is consistent with stringent $\leq$$8 \pm 2\%$ constraints on the integrated 8.44~GHz circular polarization from a previous 11-hr  observation of LSR~J1835+3259 \citep{Hallinan2008ApJ...684..644H}.  Indeed, we do not detect circular polarization in its resolved radio lobes in any epoch (Table~\ref{tab:imaging}).
For the less resolved and brighter quiescent emission in Epoch 1, our noise floor gives a 95\% confidence upper limit of $\leq$$8.8$\% and $\leq$$15.5$\% circular polarization in the left and right lobes, respectively. Synchrotron emission instead produces linear polarization \citep{Gudel2002ARAA..40..217G} that has been observed at the 20\% level for Jupiter \citep{Bolton2004}. Similar measurements would confirm a synchrotron interpretation for LSR~J1835+3259, and  higher frequency observations  can further constrain electron energies \citep{Gudel2002ARAA..40..217G}.

For synchrotron emission, we can estimate electron energies because a single electron emits at a narrow range of frequencies. Its power spectrum peaks at the critical frequency
\begin{equation}
    \nu_{\mathrm{crit}} \approx\frac{3}{2} \gamma^2 \, \nu_c \sin\alpha \,
\end{equation}
where $\alpha$ is the pitch angle \citep{Gudel2002ARAA..40..217G}. For $\nu_{\mathrm{crit}} \approx 8.5$~GHz, electrons at the centroids of our target's resolved radio lobes will have $\gamma \approx 30$. These high Lorentz factors correspond to $15$~MeV and are comparable to Jovian radiation belt electron energies up to tens of MeV \citep{Bolton2004, Kollmann2018JGRA..123.9110K}. 

\begin{table}[h]
\begin{center}
\begin{minipage}{\textwidth}
\caption{Imaging: Quiescent flux densities}\label{tab:imaging}
\begin{tabular*}{\textwidth}{@{\extracolsep{\fill}}lccccc@{\extracolsep{\fill}}}
\toprule%
Epoch                                   &
$\sigma_{\mathrm{rms}}$\footnotemark[1]  & 
$F_{\nu, \mathrm{peak}}$\footnotemark[2] & 
Integrated $F_{\nu}$\footnotemark[2]     & 
Circ. Poln.\footnotemark[3] \\
                &
($\mu$Jy/beam)  & 
($\mu$Jy/beam)  & 
($\mu$Jy)       & 
(\%)\\
\midrule
1:  Q	&	13, 12	&	\,\,271$\pm$12\,, 155$\pm$12 	&	342$\pm$25\,, 201$\pm$26	&	$\leq$8.8\,, $\leq$15.5	    \\
2:	Q	&	12, 12	&	113$\pm$11\,, $98\pm11$	        &	235$\pm$33\,, 225$\pm$35	&	$\leq$21.4\,, $\leq$25.0	\\
3:	Q	&	13, 13	&	\,\,\,\,96$\pm$12\,, 65$\pm$12	&	189$\pm$34\,, 164$\pm$40	&	$\leq$27.9\,, $\leq$44.0	\\
\botrule
\end{tabular*}
\footnotetext[1]{Quiescent emission images: total, circularly polarized (Stokes I, Stokes V)}
\footnotetext[2]{Quiescent emission: left lobe, right lobe}
\footnotetext[3]{No circularly polarized emission was detected. We give 95\% confidence upper limits for the absolute value of percent circular polarization in the peak emission calculated with $\sigma_{\mathrm{rms}}$ from the Stokes V image.} 
\end{minipage}
\end{center}
\end{table}

Relativistic electrons lose energy as they emit synchrotron radiation, giving a  cooling time in seconds \citep{Gudel2002ARAA..40..217G}
\begin{equation}
    \tau \approx \frac{6.7 \times 10^8}{B^2 \gamma}  \,.
\end{equation}
For LSR~J1835+3259,  we estimate $\tau \approx 65$~days,  yet the double-lobed structure that we observe  persists for over a year. For the left lobe, its integrated flux varies from $189\pm34$ to $342\pm25$~$\mu$Jy between all epochs, and the right lobe varies from $164\pm40$ to $225\pm35$~$\mu$Jy  (Table \ref{tab:imaging}). Although unresolved, this level of quiescent emission at the same observing frequencies was also present over a decade ago \citep{Berger2006ApJ...648..629B, Hallinan2008ApJ...684..644H, Berger2008ApJ...676.1307B}.

Flares can impulsively accelerate electrons \citep{Gudel2002ARAA..40..217G}, and ultracool dwarfs similar in spectral type to LSR~J1835+3259 such as Trappist-1 can flare at optical wavelengths as frequently as once per $\sim$day for the lowest-energy flares or once per $\sim$month for higher energy flares \citep{Paudel2019MNRAS.486.1438P}.  However, Trappist-1 lacks detectable radio emission \citep{Pineda2018ApJ...866..155P}, as do most ultracool dwarfs of similar spectral type to  LSR~J1835+3259 \citep{kao2023a}. Curiously, the flare star UV~Ceti (M5.5 spectral type) temporarily displayed a double-lobed structure at 8.4~GHz during a radio flare, but that structure disappeared within hours \cite{Benz1998AA...331..596B}. Double-lobed flares are unlikely to explain persistent quiescent emission from that star or LSR~J1835+3259.  

Instead, radiation belts have been observed for decades to maintain MeV electrons in equatorial magnetosphere regions of Solar System planets. They offer an analogy for interpreting LSR~J1835+3259's double-lobed quiescent emission. In contrast to impulsive acceleration from flares, radiation belt electrons undergo sustained adiabatic heating as they encounter stronger magnetic fields during inward radial diffusion in Jupiter's magnetosphere \citep{Kollmann2018JGRA..123.9110K}. Intriguingly, recent radiation belt modeling for massive stars can also reproduce 8.4~GHz quiescent radio luminosities from LSR~J1835+3259 \citep{Leto2021MNRAS.507.1979L}. The double-lobed and axisymmetric geometry observed from its quiescent radio emission is similar to the radio morphology of Jupiter's radiation belts \citep{Bolton2002Natur.415..987B} and consistent with a belt-like structure about the magnetic equator for an edge-on system like LSR~J1835+3259 (Figures \ref{fig:images} and \ref{fig:aurora+belt}).  Jupiter's GHz radiation belts trace its highest energy electrons and are more compact than its MHz radiation belts \citep{Bolton2002Natur.415..987B, Bolton2004, girard2016}.  97.5~GHz quiescent emission from LSR~J1835+3259 raises the possibility that its 8.5~GHz emission may similarly trace less energetic electrons at more extended distances in its magnetosphere, calling for comparisons to resolved imaging at higher frequencies.

We conclude that quiescent radio emission around LSR J1835+3259 exhibits properties consistent with an extrasolar analog to the Jovian radiation belts: its long-lived double-lobed structure (1) is morphologically similar to the Jovian radiation belts, (2) with an $\approx$18~$R_{\mathrm{UCD}}$ lobe separation  implying $\sim$MeV electrons confined in a magnetic dipole field and (3) suggesting acceleration mechanisms, distinct from flare activity, that rely on rapidly rotating magnetic dipole fields \citep{RoussosKollmann2021GMS...259..499R} to produce a belt-like structure of relativistic electrons in its magnetospheric equatorial regions. Our results support broader re-examination of the role that rotating magnetic dipoles may play for the non-thermal quiescent radio emission from massive stars \citep{Leto2021MNRAS.507.1979L} and fully convective M dwarfs \citep{Gudel1993ApJ...405L..63G}, for which growing evidence points to the prevalence of large-scale magnetospheres. 

Many open questions remain, including: what is the source of ultracool dwarf radiation belt plasma? Ongoing searches for their predicted planets and moons  \citep{Tamburo2022AJ....164..252T,  Limbach2021ApJ...918L..25L} may help demonstrate that volcanism from such companions seed ultracool dwarf magnetospheres in a manner similar to Io in Jupiter's magnetosphere  \citep{Hill1983phjm.book..353H}. Additionally, unlike Jupiter, flares on ultracool dwarfs \citep{Paudel2019MNRAS.486.1438P} may provide a seed population of electrons that are later accelerated to the high energies that we infer. Variability on days-long timescales such as what we observe for LSR~J1835+3259 are also observed from radiation belts around Jupiter and Saturn. They are attributed to changes in radial diffusion tied to solar weather \citep{Tsuchiya2011JGRA..116.9202T, Kollmann2017NatAs...1..872K}. For isolated ultracool dwarfs such as LSR~J1835+3259, we postulate that their flaring activity may similarly perturb radial diffusion of radiation belt electrons while contributing to their population. 

Beginning with the discovery of ultracool dwarf radio emission first announced in this Journal \citep{Berger2001Natur.410..338B} and the later confirmation of aurorae occurring on ultracool dwarfs \citep{Hallinan2015Natur.523..568H}, our result completes a paradigm in which planetary-type radio emissions emerge at the bottom of the stellar sequence as stellar-like flaring activity subsides.


\clearpage
\section{Methods} \label{sec:methods}

\subsection{Target parameters}
In addition to the parameter modeling discussed in the Main Text, absorption line modeling for LSR~J1835+3259 gives a higher $2800\pm30$~K effective temperature corresponding to a lower mass, young age, and inflated radius ($22\pm4$~Myr, $55 \pm 4$~$M_{\mathrm{J}}$, $2.1\pm0.1$~$R_\mathrm{J}$ \citep{Berdyugina2017ApJ...847...61B}). 
However, this higher temperature is inconsistent with its spectral type and may be subject to systematic effects in the model atmospheric spectra. Indeed, the young inferred age does not exceed typical M dwarf disk dissipation timescales \citep{Binks2017MNRAS.469..579B} and no infrared excess indicates the presence of a disk \citep{Avenhaus2012AA...548A.105A}. Furthermore, LSR~J1835+3259 does not have detectable lithium absorption in its atmosphere \citep{ReinersBasri2009ApJ...705.1416R}, indicating that its mass is likely higher than the  $\approx$65~$M_{\mathrm{J}}$ mass threshold for lithium depletion to occur and that its age is older than the depletion timescale  \cite{Basri2000ARAA..38..485B, Kirkpatrick2008ApJ...689.1295K}. Instead, the properties we adopt ($\geq$500~Myr, $77.28 \pm 10.34$~$M_{\mathrm{J}}$, $1.07 \pm 0.05$~$R_\mathrm{J}$ \citep{Filipazzo2015ApJ...810..158F}) are consistent with these multiwavelength observations of LSR~J1835+3259. We summarize our adopted target properties in Table~\ref{tab:target}.

\subsection{Observations}
The HSA combines the Very Long Baseline Array (VLBA, ten 25-m dishes), the Karl G. Jansky Very Large Array (VLA, twenty-seven 25-m dishes) as a phased array, Greenbank Telescope (GBT, single 100-m dish), and Effelsberg Telescope (EB, single 100-m dish).  Not all epochs successfully included all telescopes because of weather, equipment failures, and site closings. Additionally, LSR~J1835+3259 was visible on the longest baseline from the VLBA dish at Mauna Kea, Hawaii (MK) to EB in Bad M\"{u}nstereifel, Germany (10328~km) for no more than one hour per observation. We prioritized time on EB to increase observational sensitivity and long baselines to the telescopes on the continental USA.  Table~\ref{tab:spatial} summarizes presented HSA observations.

To incorporate the VLA in a very long baseline interferometry (VLBI) observation, all the antennas in the array must be phase corrected and summed (i.e. phased).   After this is done, the VLA can be treated as a single element in the VLBI array with a primary beam equal to the synthesized beam of the VLA. Along with the phased-VLA as a single telescope that is correlated with the other HSA telescopes, the phased-VLA data can be used as a regular VLA observation. 

We observed in A and B configurations for the VLA, giving phased-VLA primary beams of $\sim$$0.2''$ and $0.6''$ (half power beam width), respectively,  at our X-band  (8.5~GHz) observing frequency. To obtain the full sensitivity of the phased-VLA observations, the position of LSR~J1835+3259 needed to be within $0.1''$ of the center of the field when observing in A configuration and $0.3''$ of the center of the field when observing in B configuration.  

This can pose a challenge for an object with as high proper motion and parallax as LSR~J1835+3259 (Table \ref{tab:target}). \textit{Gaia} Data Release 2 measured proper and parallax motions exceeding  2~mas~day$^{-1}$ and 1~mas~day$^{-1}$, respectively \citep{Gaia20182018AA...616A...1G}. To ensure target capture, a week to a month prior to one or more HSA observations, we obtained a 60\,min observation using only the VLA in array mode at X band to image and locate the position of the target. 
									
We also observed the VLBA standard phase calibrator J1835+3241 to calibrate the phase errors from atmospheric fluctuations. This phase calibrator was within 0.33$^{\circ}$ of our target. Our phase calibration cycle periods were 4~min and 8~min for HSA and VLA-only observing blocks, respectively. During HSA observing, we also phased the VLA every 10 minutes with this same phase calibrator to maintain coherence across the VLA and observed J1848+3219 every $\sim$2~hours for fringe-finding. Finally, we observed 3C286 as a flux calibrator for the VLA in each epoch to allow for independent analyses of the phased-VLA data.

\subsection{Calibrations} \label{sec:calibrations}

For  HSA observations, we applied standard phase reference VLBI data reduction methods \citep{Torres2007ApJ...671.1813T} using the Astronomical Image Processing System \textit{AIPS} package by the National Radio Astronomy Observatory  \citep{Greisen2003ASSL..285..109G}. The target is too faint to self-calibrate, so it was phased referenced to the nearby calibrator J1835+3241. No further calibration to account for atmospheric differences between the target and the calibrator were required due to the excellent proximity of our phase calibrator (0.33$^{\circ}$ separation). 

The narrow 256~MHz bandwidth of the HSA observations and 8.4~GHz center frequency  largely avoids radio frequency interference as a source of noise. Nevertheless, we carefully examine the data from each epoch to identify and remove bad data. No data from MK was collected for Epoch 1, and calibrating the MK-EB baseline in Epochs 2 and 3 proved very difficult. This was because of the short amount of time both telescopes overlapped and the fact that they created a single very long baseline that was not similar to any other baseline in the observations.  For this reason, in Epoch 2 we did not include the MK-EB baseline and in Epoch 3 we did not include MK at all in final imaging. 

For each epoch, we calculated the combined apparent motion of our target from both proper motion and parallax. For the latter, we assumed a geocentric observer on an orbit with negligible eccentricity and take the location of the Sun as the Solar System barycenter. These assumptions give parallax offsets that are within 2\% of the true parallax amplitude \citep{Greene1985spas.book.....G} and well within our resolving power (Table~\ref{tab:spatial}) while greatly simplifying our calculations.  LSR~J1835+3259 moves $0.4-0.6$~mas in right ascension and $0.3-0.9$~mas in declination per 5-hour observation.  Our observations can resolve this apparent motion, so we compute an ``effective proper motion'' by assuming  our target takes an approximately linear path on the sky over each observation. We then correct for apparent motion smearing using the \textit{AIPS} task \texttt{clcor}.

Circular polarization, which is the difference in the right and left circularly polarized data, can distinguish between and characterize electron cyclotron maser, gyrosynchrotron, and synchrotron emissions.  Since all telescopes in the HSA use circularly polarized feeds, it is easy for even slightly incorrect amplitude calibration to produce spurious instrumental circular polarization.  To ensure that any circular polarization detected was from LSR\,J1835+3259 rather than errors in the calibration, we checked the circular polarizations of our calibrators. These showed instrumental contamination resulting in $\sim$$7-10$\% spurious circular polarization in the HSA observations.  We separately inspect the VLBA-only and VLA-only data on the phase calibrator and find that these data contain $\simeq 0.1\%$  circular polarization  from instrumental contamination and/or circular polarization intrinsic to the calibrator. To correct the amplitude calibration on the HSA, in each epoch we self-calibrated \citep{Cornwell1989ASPC....6..185C} the phase calibrator to reduce instrumental polarization to $\lesssim 1\%$. We then transferred the resulting amplitude calibrations to our target to reduce any instrumental circular polarization to the $\sim$1\% level.

\subsection{Timeseries}

\begin{figure}[t]
\centering
\includegraphics[width=1\textwidth]{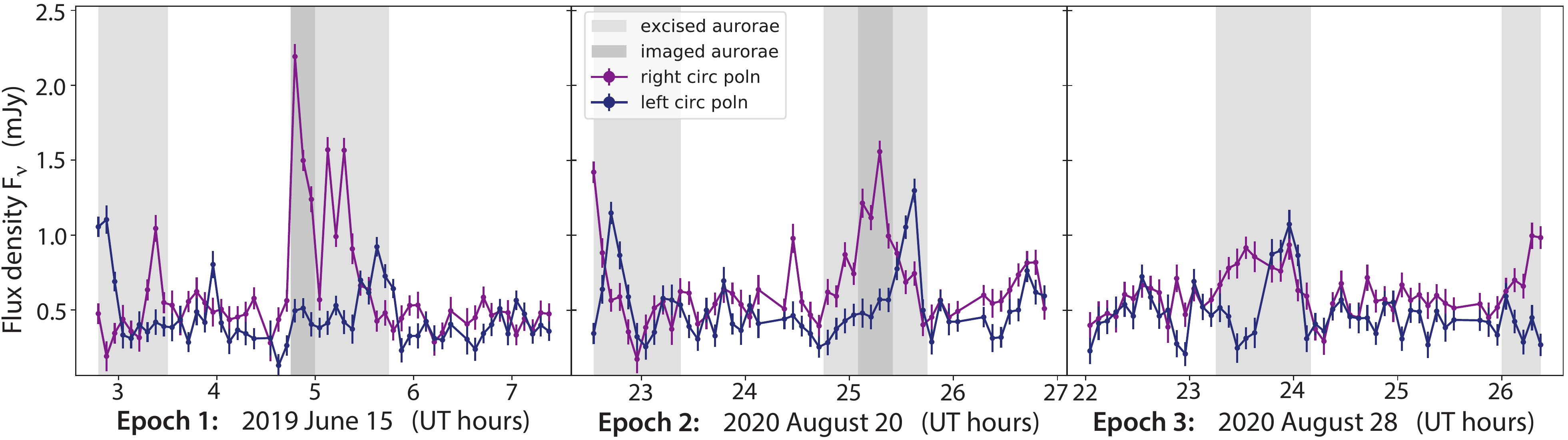}
\caption{Timeseries for LSR~J1835+3259 using only phased-VLA data from each epoch binned into 5 min intervals showing the right circularly polarized emission (magenta), left circularly polarized emission (blue) and excised aurorae (grey). Calibrator and autophasing scans that extended for $\sim$10 minutes and occurred $\sim$2 hours apart are evident in the data and partially coincided with an aurora in Epoch 3. }\label{fig:timeseries}
\end{figure}

Radio aurora on LSR~J1835+3259 manifest as bright, periodic and strongly circularly polarized bursts every $2.84 \pm 0.01$ hr \citep[e.g.][]{Hallinan2008ApJ...684..644H}
that are clearly evident even in the stand-alone phased-VLA data. LSR~J1835+3259 is unresolved in phased-VLA data, so we used this data to produce timeseries (Figure~\ref{fig:timeseries}) using the \textit{AIPS} task \texttt{dftpl}, which is specifically designed for unresolved objects. We find that two auroral bursts were partially or fully detected in each epoch for both the right and left circularly polarized data.

\subsection{Imaging: Quiescent Emission} \label{sec:NAEimaging}

To create images of the quiescent emission, we excised auroral bursts identified in the timeseries (Figure~\ref{fig:timeseries}) and imaged the remaining data in each epoch (Figure \ref{fig:images}).  All images presented in this Letter were imaged using the \textit{AIPS} task \texttt{imagr} with a 0.1\,mas pixel size to give $4-6$ pixels across the narrowest part of the synthesized beam. We used a Briggs robust weighting \citep{Briggs...1995AAS...18711202B} of 0.0, which balances between uniform and natural weighting to allow both high resolution and sensitivity to non-point sources.

We observe a double-lobed morphology in each epoch (Figure \ref{fig:images}). Detailed modeling of the quiescent emission morphology to distinguish between different morphology types is outside of the scope of this Letter.  Instead, we measure quiescent emission source sizes and flux densities using the \textit{AIPS} task \texttt{jmfit}. In each epoch image, we fit two elliptical Gaussians with freely floating centers, sizes, peaks, and integrated flux densities (Tables \ref{tab:spatial}, \ref{tab:imaging}, \ref{tab:fits}). The left lobe and right lobe are resolved along approximately the east-west axis in the later two epochs. Measured integrated flux densities are consistent with those reported in the literature \citep{Berger2001Natur.410..338B, Hallinan2008ApJ...684..644H, Berger2008ApJ...676.1307B}.

Finally, in order to help distinguish between synchrotron and gyrosynchrotron emission, we imaged the right minus the left circular polarization (total circular polarization; Stokes V) for the quiescent emission from LSR~J1835+3259 in each epoch (Table \ref{tab:imaging}).  There was no detectable circular polarization above the $12-13$~$\mu$Jy/beam noise floors of the Stokes V images. As a further check, we also image our target using data from the VLBA-only and stand-alone phased-VLA. We find no convincing circular polarization emission to an rms noise floor of 37 and 30 $\mu$Jy/beam respectively. These non-detections are consistent with low integrated circular polarization ($\sim$$8 \pm 2\%$) measured in a previous 11 hr VLA observation at 8.44~GHz that also averages over circularly polarized but periodically bursting aurorae \citep{Hallinan2008ApJ...684..644H}.
For our brightest quiescent lobe, $\sim$8\% circular polarization  would be a $\lesssim 2\sigma$ source in the circular polarization images made from the HSA data.

\subsection{Imaging: Auroral Bursts}
We imaged auroral bursts in the same way as described in Section \ref{sec:NAEimaging}. However, first we removed the quasi-steady quiescent emission. We used the \textit{AIPS} task \texttt{uvsub} to subtract a model of the quiescent emission  obtained from the images for each epoch shown in Figure \ref{fig:images}. As a check, we also subtracted this model from our quiescent-only datasets and re-imaged the data to ensure that no flux remained.

We then imaged only time ranges containing right circularly polarized auroral bursts, which are brighter in our data (Figure \ref{fig:timeseries}).  To obtain a high signal-to-noise, we imaged the brightest 15-20 minutes noted in Figure \ref{fig:timeseries}. These shorter time ranges also avoided averaging longer periods of data with very different flux densities, which can cause artifacts in interferometric imaging. 

We imaged right circularly polarized auroral bursts from Epochs 1 and 2. In Epoch 3, aurorae were too faint to confidently image, which was further exacerbated by a set of $\sim$10~min extended calibration scans coinciding with one of the auroral bursts. We also attempted to image left circularly polarized auroral bursts, but these were too faint to be imaged for all epochs. Figure \ref{fig:aurora+belt} shows the right circularly polarized auroral burst from Epoch 2 and shows that it is morphologically distinct from the quiescent radio lobes.

Finally we fit an elliptical Gaussian with the \textit{AIPS} task \texttt{jmfits} to measure the spatial extent and location of the right circularly polarized auroral bursts in Epochs 1 and 2 (Tables \ref{tab:fits} and \ref{tab:aurorae}). In Epoch 1, the auroral burst appears unresolved and associated with the inner part of the right lobe.  In Epoch 2, the burst is consistent with both being unresolved or being marginally resolved along approximately the east-west axis with a minor axis of $\sim 0.4$~mas (Figure \ref{fig:aurora+belt}).

\backmatter
\bmhead{Availability of data and materials}
All radio data are available on the National Radio Astronomy Archive under VLBA Program BK222, PI Kao.


\bmhead{Acknowledgments}
MK thanks M. Claussen for help in scheduling observations and coordinating with all arrays, and J.S. Pineda, G. Hallinan, R. Murray-Clay, and J. Fortney for input. Support was provided by NASA through the NASA Hubble Fellowship grant HST-HF2-51411.001-A awarded by the Space Telescope Science Institute, which is operated by the Association of Universities for Research in Astronomy, Inc., for NASA, under contract NAS5-26555; and by the Heising-Simons Foundation through the 51 Pegasi b Fellowship grant 2021-2943. This work is based on observations made with the National Radio Astronomy Observatory's Very Long Baseline Array (VLBA), Karl G. Jansky Very Large Array (VLA), the Green Bank Observatory's Robert C. Byrd Green Bank Telescope (GBT), and the 100-m telescope of the Max-Planck-Institut für Radioastronomie at Effelsberg. The National Radio Astronomy Observatory and the Green Bank Observatory are facilitates of the National Science Foundation operated under cooperative agreement by Associated Universities, Inc.  This work also made use of the SIMBAD and VizieR databases, operated at CDS, Strasbourg, France; and the European Space Agency (ESA) mission \textit{Gaia} (\url{https://www.cosmos.esa.int/gaia}), processed by the Gaia Data Processing and Analysis Consortium (DPAC, \url{https://www.cosmos.esa.int/web/gaia/dpac/consortium}). 

\bmhead{Author contributions}
This Letter resulted from an equal lead effort between MK and AM.  MK conceived of and led the experiment design and proposal writing, coordinated and led observations, reduced all VLA-only pre-observing blocks, collaborated on HSA data analysis, and led manuscript writing.  AM worked closely with MK to design the proposed observations, coordinate them, reduce all HSA science data presented in this Letter, and contributed to manuscript writing. JRV contributed to the proposal design and manuscript writing.  ES contributed funding to support MK during the proposal design, engaged in discussions that shaped this work, and contributed to manuscript revisions.

\bmhead{Author Information} The authors declare no competing interests.  Supplementary Information is available for this paper. Correspondence and requests or materials should be addressed to MK (mmkao@ucsc.edu). 


\clearpage
\section{Extended Data}

\begin{table}[h]
\begin{center}
\begin{minipage}{\textwidth}
\caption{Imaging: Lobe \& Aurorae Fits}\label{tab:fits}
\begin{tabular*}{\textwidth}{@{\extracolsep{\fill}}lllcc@{\extracolsep{\fill}}}
\toprule%
Epoch: Lobe\footnotemark[1]                 & 
Centroid RA             & 
Centroid Dec            &
Major axis              &
Minor axis              \\
                        &     
(J2000: hms)                 & 
(J2000: dms)                 &
(mas)                   & 
(mas)                   \\
\midrule
1:	R	         &	18 35 37.771401(2)  &	32 59 38.77150(7)  &	0.37	$\pm$	0.17	&	-- \\
\,\,\,\,\,	L    &	18 35 37.771464(1)  &	32 59 38.77217(4)  &	0.76	$\pm$	0.10	&	--	\\
\,\,\,\,\,	A    &	18 35 37.771409(3)  &	32 59 38.7714(10)  &	1.19	$\pm$	0.32	&	--  \\
\midrule
2:	R	         &	18 35 37.750313(5)  &	32 59 37.8258(10)  &	1.42	$\pm$	0.25	&	0.71	$\pm$	0.12	\\
\,\,\,\,\,	L	 &	18 35 37.750407(3)  &	32 59 37.82694(9)  &	1.23	$\pm$	0.20	&	0.58	$\pm$	0.10	\\
\,\,\,\,\,	A	 &	18 35 37.750354(4)  &	32 59 37.8263(30)  &	3.12	$\pm$	0.63	&	0.40	$\pm$	0.00	\\
\midrule
3:	R	         &	18 35 37.749064(7)  &	32 59 37.7927(20)  &	1.64	$\pm$	0.42	&	0.83	$\pm$	0.19	\\
\,\,\,\,\,	L	 &	18 35 37.749165(5)  &	32 59 37.7938(10)  &	1.34	$\pm$	0.26	&	0.66	$\pm$	0.11	\\
\botrule
\end{tabular*}
\footnotetext{See Tables \ref{tab:imaging} and \ref{tab:aurorae} for peak and integrated flux densities. Uncertainties in the least significant digit are given in parentheses for coordinates, which are for midnight in International Atomic Time on the epoch date.  Minor axes in Epoch 1 are unresolved.}
\footnotemark[1]{R: right lobe; L: left lobe; A: aurora}
\end{minipage}
\end{center}
\end{table}

\begin{table}[h]
\begin{center}
\begin{minipage}{0.6\textwidth}
\caption{Imaging: Aurorae flux densities}\label{tab:aurorae}
\begin{tabular*}{\textwidth}{@{\extracolsep{\fill}}lccc@{\extracolsep{\fill}}}
\toprule%
Epoch                                       &
$\sigma_{\mathrm{rms}}$\footnotemark[2]     & 
$F_{\nu, \mathrm{peak}}$\footnotemark[3]    & 
Integrated $F_{\nu}$\footnotemark[3]        \\
                &
($\mu$Jy/beam)  & 
($\mu$Jy/beam)  & 
($\mu$Jy)       \\
\midrule
1   &	57	&	908$\pm$69	&	965$\pm$125	\\
2   &	43	&	237$\pm$40	&	537$\pm$123	\\
3   &	43	&    $\leq$129	&	---	        \\
\botrule
\end{tabular*}
\footnotetext{All values given for right circularly polarized (RR) emission.}
\end{minipage}
\end{center}
\end{table}

\clearpage






\bibliography{biblio}

\begin{thebibliography}{10}
\expandafter\ifx\csname url\endcsname\relax
  \def\url#1{\burl{#1}}\fi
\expandafter\ifx\csname urlprefix\endcsname\relax\def\urlprefix{URL }\fi
\providecommand{\bibinfo}[2]{#2}
\providecommand{\eprint}[2][]{\url{#2}}
\providecommand{\doi}[1]{\url{https://doi.org/#1}}
\bibcommenthead

\bibitem{maukFox2010}
\bibinfo{author}{{Mauk}, B.~H.} \& \bibinfo{author}{{Fox}, N.~J.}
\newblock \bibinfo{title}{{Electron radiation belts of the solar system}}.
\newblock \emph{\bibinfo{journal}{Journal of Geophysical Research (Space
  Physics)}} \textbf{\bibinfo{volume}{115}}~(A14), \bibinfo{pages}{A12220}
  (\bibinfo{year}{2010}).
\newblock \doi{10.1029/2010JA015660} .

\bibitem{Bolton2002Natur.415..987B}
\bibinfo{author}{{Bolton}, S.~J.} \emph{et~al.}
\newblock \bibinfo{title}{{Ultra-relativistic electrons in Jupiter's radiation
  belts}}.
\newblock \emph{\bibinfo{journal}{Nature}}
  \textbf{\bibinfo{volume}{415}}~(6875), \bibinfo{pages}{987--991}
  (\bibinfo{year}{2002}).
\newblock \doi{10.1038/415987a} .

\bibitem{Bolton2004}
\bibinfo{author}{{Bolton}, S.~J.}, \bibinfo{author}{{Thorne}, R.~M.},
  \bibinfo{author}{{Bourdarie}, S.}, \bibinfo{author}{{de Pater}, I.} \&
  \bibinfo{author}{{Mauk}, B.}
\newblock \emph{\bibinfo{title}{{Jupiter's inner radiation belts}}},
  \bibinfo{pages}{671--688} (\bibinfo{year}{2004}).

\bibitem{Kollmann2018JGRA..123.9110K}
\bibinfo{author}{{Kollmann}, P.} \emph{et~al.}
\newblock \bibinfo{title}{{Electron Acceleration to MeV Energies at Jupiter and
  Saturn}}.
\newblock \emph{\bibinfo{journal}{Journal of Geophysical Research (Space
  Physics)}} \textbf{\bibinfo{volume}{123}}~(11), \bibinfo{pages}{9110--9129}
  (\bibinfo{year}{2018}).
\newblock \doi{10.1029/2018JA025665} .

\bibitem{Gudipati2021NatAs...5..276G}
\bibinfo{author}{{Gudipati}, M.~S.}, \bibinfo{author}{{Henderson}, B.~L.} \&
  \bibinfo{author}{{Bateman}, F.~B.}
\newblock \bibinfo{title}{{Laboratory predictions for the night-side surface
  ice glow of Europa}}.
\newblock \emph{\bibinfo{journal}{Nature Astronomy}}
  \textbf{\bibinfo{volume}{5}}, \bibinfo{pages}{276--282}
  (\bibinfo{year}{2021}).
\newblock \doi{10.1038/s41550-020-01248-1} .

\bibitem{Nichols2012ApJ...760...59N}
\bibinfo{author}{{Nichols}, J.~D.} \emph{et~al.}
\newblock \bibinfo{title}{{Origin of Electron Cyclotron Maser Induced Radio
  Emissions at Ultracool Dwarfs: Magnetosphere-Ionosphere Coupling Currents}}.
\newblock \emph{\bibinfo{journal}{The Astrophysical Journal}}
  \textbf{\bibinfo{volume}{760}}~(1), \bibinfo{pages}{59}
  (\bibinfo{year}{2012}).
\newblock \doi{10.1088/0004-637X/760/1/59},
  \bibinfo{eprint}{{\href{https://arxiv.org/abs/1210.1864}{{arXiv:1210.1864}}}}
   {[astro-ph.SR]}.

\bibitem{Turnpenney2017MNRAS.470.4274T}
\bibinfo{author}{{Turnpenney}, S.}, \bibinfo{author}{{Nichols}, J.~D.},
  \bibinfo{author}{{Wynn}, G.~A.} \& \bibinfo{author}{{Casewell}, S.~L.}
\newblock \bibinfo{title}{{Auroral radio emission from ultracool dwarfs: a
  Jovian model}}.
\newblock \emph{\bibinfo{journal}{Monthly Notices of the Royal Astronomical
  Society}} \textbf{\bibinfo{volume}{470}}~(4), \bibinfo{pages}{4274--4284}
  (\bibinfo{year}{2017}).
\newblock \doi{10.1093/mnras/stx1508},
  \bibinfo{eprint}{{\href{https://arxiv.org/abs/1706.04679}{{arXiv:1706.04679}}}}
   {[astro-ph.SR]}.

\bibitem{Saur2021AA...655A..75S}
\bibinfo{author}{{Saur}, J.} \emph{et~al.}
\newblock \bibinfo{title}{{Brown dwarfs as ideal candidates for detecting UV
  aurora outside the Solar System: Hubble Space Telescope observations of 2MASS
  J1237+6526}}.
\newblock \emph{\bibinfo{journal}{Astronomy \& Astrophysics}}
  \textbf{\bibinfo{volume}{655}}, \bibinfo{pages}{A75} (\bibinfo{year}{2021}).
\newblock \doi{10.1051/0004-6361/202040230},
  \bibinfo{eprint}{{\href{https://arxiv.org/abs/2109.00827}{{arXiv:2109.00827}}}}
   {[astro-ph.SR]}.

\bibitem{Hallinan2015Natur.523..568H}
\bibinfo{author}{{Hallinan}, G.} \emph{et~al.}
\newblock \bibinfo{title}{{Magnetospherically driven optical and radio aurorae
  at the end of the stellar main sequence}}.
\newblock \emph{\bibinfo{journal}{Nature}} \textbf{\bibinfo{volume}{523}},
  \bibinfo{pages}{568--571} (\bibinfo{year}{2015}).
\newblock \doi{10.1038/nature14619},
  \bibinfo{eprint}{{\href{https://arxiv.org/abs/1507.08739}{{arXiv:1507.08739}}}}
   {[astro-ph.SR]}.

\bibitem{Kao2016ApJ...818...24K}
\bibinfo{author}{{Kao}, M.~M.} \emph{et~al.}
\newblock \bibinfo{title}{{Auroral Radio Emission from Late L and T Dwarfs: A
  New Constraint on Dynamo Theory in the Substellar Regime}}.
\newblock \emph{\bibinfo{journal}{The Astrophysical Journal}}
  \textbf{\bibinfo{volume}{818}}, \bibinfo{pages}{24} (\bibinfo{year}{2016}).
\newblock \doi{10.3847/0004-637X/818/1/24},
  \bibinfo{eprint}{{\href{https://arxiv.org/abs/1511.03661}{{arXiv:1511.03661}}}}
   {[astro-ph.SR]}.

\bibitem{Pineda2017ApJ...846...75P}
\bibinfo{author}{{Pineda}, J.~S.}, \bibinfo{author}{{Hallinan}, G.} \&
  \bibinfo{author}{{Kao}, M.~M.}
\newblock \bibinfo{title}{{A Panchromatic View of Brown Dwarf Aurorae}}.
\newblock \emph{\bibinfo{journal}{The Astrophysical Journal}}
  \textbf{\bibinfo{volume}{846}}, \bibinfo{pages}{75} (\bibinfo{year}{2017}).
\newblock \doi{10.3847/1538-4357/aa8596},
  \bibinfo{eprint}{{\href{https://arxiv.org/abs/1708.02942}{{arXiv:1708.02942}}}}
   {[astro-ph.SR]}.

\bibitem{Williams2014ApJ...785....9W}
\bibinfo{author}{{Williams}, P.~K.~G.}, \bibinfo{author}{{Cook}, B.~A.} \&
  \bibinfo{author}{{Berger}, E.}
\newblock \bibinfo{title}{{Trends in Ultracool Dwarf Magnetism. I. X-Ray
  Suppression and Radio Enhancement}}.
\newblock \emph{\bibinfo{journal}{The Astrophysical Journal}}
  \textbf{\bibinfo{volume}{785}}, \bibinfo{pages}{9} (\bibinfo{year}{2014}).
\newblock \doi{10.1088/0004-637X/785/1/9},
  \bibinfo{eprint}{{\href{https://arxiv.org/abs/1310.6757}{{arXiv:1310.6757}}}}
   {[astro-ph.SR]}.

\bibitem{Hallinan2006ApJ...653..690H}
\bibinfo{author}{{Hallinan}, G.} \emph{et~al.}
\newblock \bibinfo{title}{{Rotational Modulation of the Radio Emission from the
  M9 Dwarf TVLM 513-46546: Broadband Coherent Emission at the Substellar
  Boundary?}}
\newblock \emph{\bibinfo{journal}{The Astrophysical Journal}}
  \textbf{\bibinfo{volume}{653}}, \bibinfo{pages}{690--699}
  (\bibinfo{year}{2006}).
\newblock \doi{10.1086/508678},
  \bibinfo{eprint}{{\href{https://arxiv.org/abs/astro-ph/0608556}{{astro-ph/0608556}}}}
  .

\bibitem{Kao2019MNRAS.487.1994K}
\bibinfo{author}{{Kao}, M.~M.}, \bibinfo{author}{{Hallinan}, G.} \&
  \bibinfo{author}{{Pineda}, J.~S.}
\newblock \bibinfo{title}{{Constraints on magnetospheric radio emission from Y
  dwarfs}}.
\newblock \emph{\bibinfo{journal}{Monthly Notices of the Royal Astronomical
  Society}} \textbf{\bibinfo{volume}{487}}~(2), \bibinfo{pages}{1994--2004}
  (\bibinfo{year}{2019}).
\newblock \doi{10.1093/mnras/stz1372} .

\bibitem{Leto2021MNRAS.507.1979L}
\bibinfo{author}{{Leto}, P.} \emph{et~al.}
\newblock \bibinfo{title}{{A scaling relationship for non-thermal radio
  emission from ordered magnetospheres: from the top of the main sequence to
  planets}}.
\newblock \emph{\bibinfo{journal}{Monthly Notices of the Royal Astronomical
  Society}} \textbf{\bibinfo{volume}{507}}~(2), \bibinfo{pages}{1979--1998}
  (\bibinfo{year}{2021}).
\newblock \doi{10.1093/mnras/stab2168},
  \bibinfo{eprint}{{\href{https://arxiv.org/abs/2107.11995}{{arXiv:2107.11995}}}}
   {[astro-ph.SR]}.

\bibitem{Climent2022AA...660A..65C}
\bibinfo{author}{{Climent}, J.~B.} \emph{et~al.}
\newblock \bibinfo{title}{{Radio emission in a nearby, ultra-cool dwarf binary:
  A multifrequency study}}.
\newblock \emph{\bibinfo{journal}{Astronomy \& Astrophysics}}
  \textbf{\bibinfo{volume}{660}}, \bibinfo{pages}{A65} (\bibinfo{year}{2022}).
\newblock \doi{10.1051/0004-6361/202142260} .

\bibitem{Gudel1993ApJ...405L..63G}
\bibinfo{author}{{G{\"u}del}, M.} \& \bibinfo{author}{{Benz}, A.~O.}
\newblock \bibinfo{title}{{X-Ray/Microwave Relation of Different Types of
  Active Stars}}.
\newblock \emph{\bibinfo{journal}{The Astrophysical Journal Letters}}
  \textbf{\bibinfo{volume}{405}}, \bibinfo{pages}{L63} (\bibinfo{year}{1993}).
\newblock \doi{10.1086/186766} .

\bibitem{Benz1994A&A...285..621B}
\bibinfo{author}{{Benz}, A.~O.} \& \bibinfo{author}{{Guedel}, M.}
\newblock \bibinfo{title}{{X-ray/microwave ratio of flares and coronae}}.
\newblock \emph{\bibinfo{journal}{Astronomy \& Astrophysics}}
  \textbf{\bibinfo{volume}{285}}, \bibinfo{pages}{621--630}
  (\bibinfo{year}{1994}) .

\bibitem{dePater1982}
\bibinfo{author}{{de Pater}, I.}, \bibinfo{author}{Kenderdine, S.} \&
  \bibinfo{author}{Dickel, J.~R.}
\newblock \bibinfo{title}{Comparison of the thermal and nonthermal radiation
  characteristics of jupiter at 6, 11, and 21 cm with model calculations}.
\newblock \emph{\bibinfo{journal}{Icarus}} \textbf{\bibinfo{volume}{51}}~(1),
  \bibinfo{pages}{25--38} (\bibinfo{year}{1982}).
\newblock
  \urlprefix\url{https://www.sciencedirect.com/science/article/pii/0019103582900276}.
\newblock \doi{https://doi.org/10.1016/0019-1035(82)90027-6} .

\bibitem{dePater2016}
\bibinfo{author}{de~Pater, I.}, \bibinfo{author}{Sault, R.~J.},
  \bibinfo{author}{Butler, B.}, \bibinfo{author}{DeBoer, D.} \&
  \bibinfo{author}{Wong, M.~H.}
\newblock \bibinfo{title}{Peering through jupiter\&\#x2019;s clouds with radio
  spectral imaging}.
\newblock \emph{\bibinfo{journal}{Science}}
  \textbf{\bibinfo{volume}{352}}~(6290), \bibinfo{pages}{1198--1201}
  (\bibinfo{year}{2016}).
\newblock
  \urlprefix\url{https://www.science.org/doi/abs/10.1126/science.aaf2210}.
\newblock \doi{10.1126/science.aaf2210},
  \bibinfo{eprint}{{\href{https://arxiv.org/abs/https://www.science.org/doi/pdf/10.1126/science.aaf2210}{{https://www.science.org/doi/pdf/10.1126/science.aaf2210}}}}
  .

\bibitem{BurkeFranklin1955JGR....60..213B}
\bibinfo{author}{{Burke}, B.~F.} \& \bibinfo{author}{{Franklin}, K.~L.}
\newblock \bibinfo{title}{{Observations of a Variable Radio Source Associated
  with the Planet Jupiter}}.
\newblock \emph{\bibinfo{journal}{Journal of Geophysical Research}}
  \textbf{\bibinfo{volume}{60}}~(2), \bibinfo{pages}{213--217}
  (\bibinfo{year}{1955}).
\newblock \doi{10.1029/JZ060i002p00213} .

\bibitem{Zarka1998JGR...10320159Z}
\bibinfo{author}{{Zarka}, P.}
\newblock \bibinfo{title}{{Auroral radio emissions at the outer planets:
  Observations and theories}}.
\newblock \emph{\bibinfo{journal}{Journal of Geophysical Research}}
  \textbf{\bibinfo{volume}{103}}~(E9), \bibinfo{pages}{20159--20194}
  (\bibinfo{year}{1998}).
\newblock \doi{10.1029/98JE01323} .

\bibitem{Osten2009ApJ...700.1750O}
\bibinfo{author}{{Osten}, R.~A.}, \bibinfo{author}{{Phan-Bao}, N.},
  \bibinfo{author}{{Hawley}, S.~L.}, \bibinfo{author}{{Reid}, I.~N.} \&
  \bibinfo{author}{{Ojha}, R.}
\newblock \bibinfo{title}{{Steady and Transient Radio Emission from Ultracool
  Dwarfs}}.
\newblock \emph{\bibinfo{journal}{The Astrophysical Journal}}
  \textbf{\bibinfo{volume}{700}}, \bibinfo{pages}{1750--1758}
  (\bibinfo{year}{2009}).
\newblock \doi{10.1088/0004-637X/700/2/1750},
  \bibinfo{eprint}{{\href{https://arxiv.org/abs/0905.4197}{{arXiv:0905.4197}}}}
   {[astro-ph.SR]}.

\bibitem{Williams2013ApJ...767L..30W}
\bibinfo{author}{{Williams}, P.~K.~G.}, \bibinfo{author}{{Berger}, E.} \&
  \bibinfo{author}{{Zauderer}, B.~A.}
\newblock \bibinfo{title}{{Quasi-quiescent Radio Emission from the First
  Radio-emitting T Dwarf}}.
\newblock \emph{\bibinfo{journal}{The Astrophysical Journal Letters}}
  \textbf{\bibinfo{volume}{767}}, \bibinfo{pages}{L30} (\bibinfo{year}{2013}).
\newblock \doi{10.1088/2041-8205/767/2/L30},
  \bibinfo{eprint}{{\href{https://arxiv.org/abs/1301.2321}{{arXiv:1301.2321}}}}
   {[astro-ph.SR]}.

\bibitem{kao2023a}
\bibinfo{author}{{Kao}, M.~M.} \& \bibinfo{author}{{Shkolnik }, E.}
\newblock \bibinfo{title}{{The occurrence rate of quiescent radio emission for
  ultracool dwarfs using a generalized analytical Bayesian framework}}.
\newblock \emph{\bibinfo{journal}{Monthly Notices of the Royal Astronomical
  Society}}  (\bibinfo{year}{submitted}) .

\bibitem{Richey-Yowell2020}
\bibinfo{author}{{Richey-Yowell}, T.}, \bibinfo{author}{{Kao}, M.~M.},
  \bibinfo{author}{{Pineda}, J.~S.}, \bibinfo{author}{{Shkolnik}, E.~L.} \&
  \bibinfo{author}{{Hallinan}, G.}
\newblock \bibinfo{title}{{On the Correlation between L Dwarf Optical and
  Infrared Variability and Radio Aurorae}}.
\newblock \emph{\bibinfo{journal}{The Astrophysical Journal}}
  \textbf{\bibinfo{volume}{903}}~(1), \bibinfo{pages}{74}
  (\bibinfo{year}{2020}).
\newblock \doi{10.3847/1538-4357/abb826},
  \bibinfo{eprint}{{\href{https://arxiv.org/abs/2009.05590}{{arXiv:2009.05590}}}}
   {[astro-ph.SR]}.

\bibitem{Forbrich2009ApJ...706L.205F}
\bibinfo{author}{{Forbrich}, J.} \& \bibinfo{author}{{Berger}, E.}
\newblock \bibinfo{title}{{The First VLBI Detection of an Ultracool Dwarf:
  Implications for the Detectability of Sub-Stellar Companions}}.
\newblock \emph{\bibinfo{journal}{The Astrophysical Journal Letters}}
  \textbf{\bibinfo{volume}{706}}~(2), \bibinfo{pages}{L205--L209}
  (\bibinfo{year}{2009}).
\newblock \doi{10.1088/0004-637X/706/2/L205},
  \bibinfo{eprint}{{\href{https://arxiv.org/abs/0910.1349}{{arXiv:0910.1349}}}}
   {[astro-ph.SR]}.

\bibitem{Gaia2020yCat.1350....0G}
\bibinfo{author}{{Gaia Collaboration}}.
\newblock \bibinfo{title}{{VizieR Online Data Catalog: Gaia EDR3 (Gaia
  Collaboration, 2020)}}.
\newblock \emph{\bibinfo{journal}{VizieR Online Data Catalog}}
  \bibinfo{pages}{I/350} (\bibinfo{year}{2020}) .

\bibitem{Deshpande2012AJ....144...99D}
\bibinfo{author}{{Deshpande}, R.} \emph{et~al.}
\newblock \bibinfo{title}{{Intermediate Resolution Near-infrared Spectroscopy
  of 36 Late M Dwarfs}}.
\newblock \emph{\bibinfo{journal}{The Astronomical Journal}}
  \textbf{\bibinfo{volume}{144}}~(4), \bibinfo{pages}{99}
  (\bibinfo{year}{2012}).
\newblock \doi{10.1088/0004-6256/144/4/99},
  \bibinfo{eprint}{{\href{https://arxiv.org/abs/1207.2781}{{arXiv:1207.2781}}}}
   {[astro-ph.SR]}.

\bibitem{Filipazzo2015ApJ...810..158F}
\bibinfo{author}{{Filippazzo}, J.~C.} \emph{et~al.}
\newblock \bibinfo{title}{{Fundamental Parameters and Spectral Energy
  Distributions of Young and Field Age Objects with Masses Spanning the Stellar
  to Planetary Regime}}.
\newblock \emph{\bibinfo{journal}{The Astrophysical Journal}}
  \textbf{\bibinfo{volume}{810}}~(2), \bibinfo{pages}{158}
  (\bibinfo{year}{2015}).
\newblock \doi{10.1088/0004-637X/810/2/158},
  \bibinfo{eprint}{{\href{https://arxiv.org/abs/1508.01767}{{arXiv:1508.01767}}}}
   {[astro-ph.SR]}.

\bibitem{Harding2013ApJ...779..101H}
\bibinfo{author}{{Harding}, L.~K.} \emph{et~al.}
\newblock \bibinfo{title}{{Periodic Optical Variability of Radio-detected
  Ultracool Dwarfs}}.
\newblock \emph{\bibinfo{journal}{The Astrophysical Journal}}
  \textbf{\bibinfo{volume}{779}}~(2), \bibinfo{pages}{101}
  (\bibinfo{year}{2013}).
\newblock \doi{10.1088/0004-637X/779/2/101},
  \bibinfo{eprint}{{\href{https://arxiv.org/abs/1310.1367}{{arXiv:1310.1367}}}}
   {[astro-ph.SR]}.

\bibitem{Berger2008ApJ...676.1307B}
\bibinfo{author}{{Berger}, E.} \emph{et~al.}
\newblock \bibinfo{title}{{Simultaneous Multiwavelength Observations of
  Magnetic Activity in Ultracool Dwarfs. II. Mixed Trends in VB 10 and LSR
  1835+32 and the Possible Role of Rotation}}.
\newblock \emph{\bibinfo{journal}{The Astrophysical Journal}}
  \textbf{\bibinfo{volume}{676}}, \bibinfo{pages}{1307--1318}
  (\bibinfo{year}{2008}).
\newblock \doi{10.1086/529131},
  \bibinfo{eprint}{{\href{https://arxiv.org/abs/0710.3383}{{arXiv:0710.3383}}}}
  .

\bibitem{Hallinan2008ApJ...684..644H}
\bibinfo{author}{{Hallinan}, G.} \emph{et~al.}
\newblock \bibinfo{title}{{Confirmation of the Electron Cyclotron Maser
  Instability as the Dominant Source of Radio Emission from Very Low Mass Stars
  and Brown Dwarfs}}.
\newblock \emph{\bibinfo{journal}{The Astrophysical Journal}}
  \textbf{\bibinfo{volume}{684}}, \bibinfo{pages}{644--653}
  (\bibinfo{year}{2008}).
\newblock \doi{10.1086/590360},
  \bibinfo{eprint}{{\href{https://arxiv.org/abs/0805.4010}{{arXiv:0805.4010}}}}
  .

\bibitem{Berger2006ApJ...648..629B}
\bibinfo{author}{{Berger}, E.}
\newblock \bibinfo{title}{{Radio Observations of a Large Sample of Late M, L,
  and T Dwarfs: The Distribution of Magnetic Field Strengths}}.
\newblock \emph{\bibinfo{journal}{The Astrophysical Journal}}
  \textbf{\bibinfo{volume}{648}}, \bibinfo{pages}{629--636}
  (\bibinfo{year}{2006}).
\newblock \doi{10.1086/505787},
  \bibinfo{eprint}{{\href{https://arxiv.org/abs/astro-ph/0603176}{{astro-ph/0603176}}}}
  .

\bibitem{hughes2021AJ....162...43H}
\bibinfo{author}{{Hughes}, A.~G.}, \bibinfo{author}{{Boley}, A.~C.},
  \bibinfo{author}{{Osten}, R.~A.}, \bibinfo{author}{{White}, J.~A.} \&
  \bibinfo{author}{{Leacock}, M.}
\newblock \bibinfo{title}{{Unlocking the Origins of Ultracool Dwarf Radio
  Emission}}.
\newblock \emph{\bibinfo{journal}{The Astronomical Journal}}
  \textbf{\bibinfo{volume}{162}}~(2), \bibinfo{pages}{43}
  (\bibinfo{year}{2021}).
\newblock \doi{10.3847/1538-3881/ac02c3} .

\bibitem{Berdyugina2017ApJ...847...61B}
\bibinfo{author}{{Berdyugina}, S.~V.} \emph{et~al.}
\newblock \bibinfo{title}{{First Detection of a Strong Magnetic Field on a
  Bursty Brown Dwarf: Puzzle Solved}}.
\newblock \emph{\bibinfo{journal}{The Astrophysical Journal}}
  \textbf{\bibinfo{volume}{847}}~(1), \bibinfo{pages}{61}
  (\bibinfo{year}{2017}).
\newblock \doi{10.3847/1538-4357/aa866b},
  \bibinfo{eprint}{{\href{https://arxiv.org/abs/1709.02861}{{arXiv:1709.02861}}}}
   {[astro-ph.SR]}.

\bibitem{Avenhaus2012AA...548A.105A}
\bibinfo{author}{{Avenhaus}, H.}, \bibinfo{author}{{Schmid}, H.~M.} \&
  \bibinfo{author}{{Meyer}, M.~R.}
\newblock \bibinfo{title}{{The nearby population of M-dwarfs with WISE: a
  search for warm circumstellar dust}}.
\newblock \emph{\bibinfo{journal}{Astronomy \& Astrophysics}}
  \textbf{\bibinfo{volume}{548}}, \bibinfo{pages}{A105} (\bibinfo{year}{2012}).
\newblock \doi{10.1051/0004-6361/201219783},
  \bibinfo{eprint}{{\href{https://arxiv.org/abs/1209.0678}{{arXiv:1209.0678}}}}
   {[astro-ph.SR]}.

\bibitem{Kuzmychov2017ApJ...847...60K}
\bibinfo{author}{{Kuzmychov}, O.}, \bibinfo{author}{{Berdyugina}, S.~V.} \&
  \bibinfo{author}{{Harrington}, D.~M.}
\newblock \bibinfo{title}{{First Spectropolarimetric Measurement of a Brown
  Dwarf Magnetic Field in Molecular Bands}}.
\newblock \emph{\bibinfo{journal}{The Astrophysical Journal}}
  \textbf{\bibinfo{volume}{847}}~(1), \bibinfo{pages}{60}
  (\bibinfo{year}{2017}).
\newblock \doi{10.3847/1538-4357/aa705a},
  \bibinfo{eprint}{{\href{https://arxiv.org/abs/1705.01590}{{arXiv:1705.01590}}}}
   {[astro-ph.SR]}.

\bibitem{RoussosKollmann2021GMS...259..499R}
\bibinfo{author}{{Roussos}, E.} \& \bibinfo{author}{{Kollmann}, P.}
\newblock \bibinfo{editor}{{Maggiolo}, R.}, \bibinfo{editor}{{Andr{\'e}}, N.},
  \bibinfo{editor}{{Hasegawa}, H.} \& \bibinfo{editor}{{Welling}, D.~T.} (eds)
  \emph{\bibinfo{title}{{The Radiation Belts of Jupiter and Saturn}}}.
\newblock (eds \bibinfo{editor}{{Maggiolo}, R.}, \bibinfo{editor}{{Andr{\'e}},
  N.}, \bibinfo{editor}{{Hasegawa}, H.} \& \bibinfo{editor}{{Welling}, D.~T.})
  \emph{\bibinfo{booktitle}{Magnetospheres in the Solar System}},
  Vol.~\bibinfo{volume}{2}, \bibinfo{pages}{499} (\bibinfo{year}{2021}).
\newblock \eprint{2006.14682}.

\bibitem{Gudel2002ARAA..40..217G}
\bibinfo{author}{{G{\"u}del}, M.}
\newblock \bibinfo{title}{{Stellar Radio Astronomy: Probing Stellar Atmospheres
  from Protostars to Giants}}.
\newblock \emph{\bibinfo{journal}{Annual Review of Astronomy \& Astrophysics}}
  \textbf{\bibinfo{volume}{40}}, \bibinfo{pages}{217--261}
  (\bibinfo{year}{2002}).
\newblock \doi{10.1146/annurev.astro.40.060401.093806},
  \bibinfo{eprint}{{\href{https://arxiv.org/abs/astro-ph/0206436}{{arXiv:astro-ph/0206436}}}}
   {[astro-ph]}.

\bibitem{Paudel2019MNRAS.486.1438P}
\bibinfo{author}{{Paudel}, R.~R.} \emph{et~al.}
\newblock \bibinfo{title}{{K2 Ultracool Dwarfs Survey - V. High superflare
  rates on rapidly rotating late-M dwarfs}}.
\newblock \emph{\bibinfo{journal}{Monthly Notices of the Royal Astronomical
  Society}} \textbf{\bibinfo{volume}{486}}~(1), \bibinfo{pages}{1438--1447}
  (\bibinfo{year}{2019}).
\newblock \doi{10.1093/mnras/stz886},
  \bibinfo{eprint}{{\href{https://arxiv.org/abs/1812.07631}{{arXiv:1812.07631}}}}
   {[astro-ph.SR]}.

\bibitem{Pineda2018ApJ...866..155P}
\bibinfo{author}{{Pineda}, J.~S.} \& \bibinfo{author}{{Hallinan}, G.}
\newblock \bibinfo{title}{{A Deep Radio Limit for the TRAPPIST-1 System}}.
\newblock \emph{\bibinfo{journal}{The Astrophysical Journal}}
  \textbf{\bibinfo{volume}{866}}~(2), \bibinfo{pages}{155}
  (\bibinfo{year}{2018}).
\newblock \doi{10.3847/1538-4357/aae078},
  \bibinfo{eprint}{{\href{https://arxiv.org/abs/1806.00480}{{arXiv:1806.00480}}}}
   {[astro-ph.SR]}.

\bibitem{Benz1998AA...331..596B}
\bibinfo{author}{{Benz}, A.~O.}, \bibinfo{author}{{Conway}, J.} \&
  \bibinfo{author}{{Gudel}, M.}
\newblock \bibinfo{title}{{First VLBI images of a main-sequence star}}.
\newblock \emph{\bibinfo{journal}{Astronomy \& Astrophysics}}
  \textbf{\bibinfo{volume}{331}}, \bibinfo{pages}{596--600}
  (\bibinfo{year}{1998}) .

\bibitem{girard2016}
\bibinfo{author}{{Girard}, J.~N.} \emph{et~al.}
\newblock \bibinfo{title}{{Imaging Jupiter's radiation belts down to 127 MHz
  with LOFAR}}.
\newblock \emph{\bibinfo{journal}{Astronomy \& Astrophysics}}
  \textbf{\bibinfo{volume}{587}}, \bibinfo{pages}{A3} (\bibinfo{year}{2016}).
\newblock \doi{10.1051/0004-6361/201527518},
  \bibinfo{eprint}{{\href{https://arxiv.org/abs/1511.09118}{{arXiv:1511.09118}}}}
   {[astro-ph.EP]}.

\bibitem{Tamburo2022AJ....164..252T}
\bibinfo{author}{{Tamburo}, P.} \emph{et~al.}
\newblock \bibinfo{title}{{The Perkins INfrared Exosatellite Survey (PINES).
  II. Transit Candidates and Implications for Planet Occurrence around L and T
  Dwarfs}}.
\newblock \emph{\bibinfo{journal}{The Astronomical Journal}}
  \textbf{\bibinfo{volume}{164}}~(6), \bibinfo{pages}{252}
  (\bibinfo{year}{2022}).
\newblock \doi{10.3847/1538-3881/ac9a52},
  \bibinfo{eprint}{{\href{https://arxiv.org/abs/2210.04462}{{arXiv:2210.04462}}}}
   {[astro-ph.EP]}.

\bibitem{Limbach2021ApJ...918L..25L}
\bibinfo{author}{{Limbach}, M.~A.} \emph{et~al.}
\newblock \bibinfo{title}{{On the Detection of Exomoons Transiting Isolated
  Planetary-mass Objects}}.
\newblock \emph{\bibinfo{journal}{The Astrophysical Journal Letters}}
  \textbf{\bibinfo{volume}{918}}~(2), \bibinfo{pages}{L25}
  (\bibinfo{year}{2021}).
\newblock \doi{10.3847/2041-8213/ac1e2d},
  \bibinfo{eprint}{{\href{https://arxiv.org/abs/2108.08323}{{arXiv:2108.08323}}}}
   {[astro-ph.EP]}.

\bibitem{Hill1983phjm.book..353H}
\bibinfo{author}{{Hill}, T.~W.}, \bibinfo{author}{{Dessler}, A.~J.} \&
  \bibinfo{author}{{Goertz}, C.~K.}
\newblock \bibinfo{title}{ in \textit{{Physics of the Jovian magnetosphere. 10.
  Magnetospheric models.}}}  \bibinfo{pages}{353--394} (\bibinfo{year}{1983}).

\bibitem{Tsuchiya2011JGRA..116.9202T}
\bibinfo{author}{{Tsuchiya}, F.}, \bibinfo{author}{{Misawa}, H.},
  \bibinfo{author}{{Imai}, K.} \& \bibinfo{author}{{Morioka}, A.}
\newblock \bibinfo{title}{{Short-term changes in Jupiter's synchrotron
  radiation at 325 MHz: Enhanced radial diffusion in Jupiter's radiation belt
  driven by solar UV/EUV heating}}.
\newblock \emph{\bibinfo{journal}{Journal of Geophysical Research (Space
  Physics)}} \textbf{\bibinfo{volume}{116}}~(A9), \bibinfo{pages}{A09202}
  (\bibinfo{year}{2011}).
\newblock \doi{10.1029/2010JA016303} .

\bibitem{Kollmann2017NatAs...1..872K}
\bibinfo{author}{{Kollmann}, P.}, \bibinfo{author}{{Roussos}, E.},
  \bibinfo{author}{{Kotova}, A.}, \bibinfo{author}{{Paranicas}, C.} \&
  \bibinfo{author}{{Krupp}, N.}
\newblock \bibinfo{title}{{The evolution of Saturn's radiation belts modulated
  by changes in radial diffusion}}.
\newblock \emph{\bibinfo{journal}{Nature Astronomy}}
  \textbf{\bibinfo{volume}{1}}, \bibinfo{pages}{872--877}
  (\bibinfo{year}{2017}).
\newblock \doi{10.1038/s41550-017-0287-x} .

\bibitem{Berger2001Natur.410..338B}
\bibinfo{author}{{Berger}, E.} \emph{et~al.}
\newblock \bibinfo{title}{{Discovery of radio emission from the brown dwarf
  LP944-20}}.
\newblock \emph{\bibinfo{journal}{Nature}} \textbf{\bibinfo{volume}{410}},
  \bibinfo{pages}{338--340} (\bibinfo{year}{2001}).
\newblock
  \bibinfo{eprint}{{\href{https://arxiv.org/abs/astro-ph/0102301}{{astro-ph/0102301}}}}
  .

\bibitem{Binks2017MNRAS.469..579B}
\bibinfo{author}{{Binks}, A.~S.} \& \bibinfo{author}{{Jeffries}, R.~D.}
\newblock \bibinfo{title}{{A WISE-based search for debris discs amongst M
  dwarfs in nearby, young, moving groups}}.
\newblock \emph{\bibinfo{journal}{Monthly Notices of the Royal Astronomical
  Society}} \textbf{\bibinfo{volume}{469}}~(1), \bibinfo{pages}{579--593}
  (\bibinfo{year}{2017}).
\newblock \doi{10.1093/mnras/stx838},
  \bibinfo{eprint}{{\href{https://arxiv.org/abs/1611.07416}{{arXiv:1611.07416}}}}
   {[astro-ph.SR]}.

\bibitem{ReinersBasri2009ApJ...705.1416R}
\bibinfo{author}{{Reiners}, A.} \& \bibinfo{author}{{Basri}, G.}
\newblock \bibinfo{title}{{A Volume-Limited Sample of 63 M7-M9.5 Dwarfs. I.
  Space Motion, Kinematic Age, and Lithium}}.
\newblock \emph{\bibinfo{journal}{The Astrophysical Journal}}
  \textbf{\bibinfo{volume}{705}}~(2), \bibinfo{pages}{1416--1424}
  (\bibinfo{year}{2009}).
\newblock \doi{10.1088/0004-637X/705/2/1416},
  \bibinfo{eprint}{{\href{https://arxiv.org/abs/0909.4647}{{arXiv:0909.4647}}}}
   {[astro-ph.SR]}.

\bibitem{Basri2000ARAA..38..485B}
\bibinfo{author}{{Basri}, G.}
\newblock \bibinfo{title}{{Observations of Brown Dwarfs}}.
\newblock \emph{\bibinfo{journal}{Annual Review of Astronomy \& Astrophysics}}
  \textbf{\bibinfo{volume}{38}}, \bibinfo{pages}{485--519}
  (\bibinfo{year}{2000}).
\newblock \doi{10.1146/annurev.astro.38.1.485} .

\bibitem{Kirkpatrick2008ApJ...689.1295K}
\bibinfo{author}{{Kirkpatrick}, J.~D.} \emph{et~al.}
\newblock \bibinfo{title}{{A Sample of Very Young Field L Dwarfs and
  Implications for the Brown Dwarf ``Lithium Test'' at Early Ages}}.
\newblock \emph{\bibinfo{journal}{The Astrophysical Journal}}
  \textbf{\bibinfo{volume}{689}}~(2), \bibinfo{pages}{1295--1326}
  (\bibinfo{year}{2008}).
\newblock \doi{10.1086/592768},
  \bibinfo{eprint}{{\href{https://arxiv.org/abs/0808.3153}{{arXiv:0808.3153}}}}
   {[astro-ph]}.

\bibitem{Gaia20182018AA...616A...1G}
\bibinfo{author}{{Gaia Collaboration}} \emph{et~al.}
\newblock \bibinfo{title}{{Gaia Data Release 2. Summary of the contents and
  survey properties}}.
\newblock \emph{\bibinfo{journal}{Astronomy \& Astrophysics}}
  \textbf{\bibinfo{volume}{616}}, \bibinfo{pages}{A1} (\bibinfo{year}{2018}).
\newblock \doi{10.1051/0004-6361/201833051},
  \bibinfo{eprint}{{\href{https://arxiv.org/abs/1804.09365}{{arXiv:1804.09365}}}}
   {[astro-ph.GA]}.

\bibitem{Torres2007ApJ...671.1813T}
\bibinfo{author}{{Torres}, R.~M.}, \bibinfo{author}{{Loinard}, L.},
  \bibinfo{author}{{Mioduszewski}, A.~J.} \& \bibinfo{author}{{Rodr{\'\i}guez},
  L.~F.}
\newblock \bibinfo{title}{{VLBA Determination of the Distance to Nearby
  Star-forming Regions. II. Hubble 4 and HDE 283572 in Taurus}}.
\newblock \emph{\bibinfo{journal}{The Astrophysical Journal}}
  \textbf{\bibinfo{volume}{671}}~(2), \bibinfo{pages}{1813--1819}
  (\bibinfo{year}{2007}).
\newblock \doi{10.1086/522924},
  \bibinfo{eprint}{{\href{https://arxiv.org/abs/0708.4403}{{arXiv:0708.4403}}}}
   {[astro-ph]}.

\bibitem{Greisen2003ASSL..285..109G}
\bibinfo{author}{{Greisen}, E.~W.}
\newblock \bibinfo{editor}{{Heck}, A.} (ed.) \emph{\bibinfo{title}{{AIPS, the
  VLA, and the VLBA}}}.
\newblock (ed.\bibinfo{editor}{{Heck}, A.})
  \emph{\bibinfo{booktitle}{Information Handling in Astronomy - Historical
  Vistas}}, Vol. \bibinfo{volume}{285} of \emph{\bibinfo{series}{Astrophysics
  and Space Science Library}}, \bibinfo{pages}{109} (\bibinfo{year}{2003}).

\bibitem{Greene1985spas.book.....G}
\bibinfo{author}{{Green}, R.~M.}
\newblock \emph{\bibinfo{title}{{Spherical Astronomy}}}
  (\bibinfo{year}{1985}).

\bibitem{Cornwell1989ASPC....6..185C}
\bibinfo{author}{{Cornwell}, T.} \& \bibinfo{author}{{Fomalont}, E.~B.}
\newblock \bibinfo{editor}{{Perley}, R.~A.}, \bibinfo{editor}{{Schwab}, F.~R.}
  \& \bibinfo{editor}{{Bridle}, A.~H.} (eds)
  \emph{\bibinfo{title}{{Self-Calibration}}}.
\newblock (eds \bibinfo{editor}{{Perley}, R.~A.}, \bibinfo{editor}{{Schwab},
  F.~R.} \& \bibinfo{editor}{{Bridle}, A.~H.})
  \emph{\bibinfo{booktitle}{Synthesis Imaging in Radio Astronomy}},
  Vol.~\bibinfo{volume}{6} of \emph{\bibinfo{series}{Astronomical Society of
  the Pacific Conference Series}}, \bibinfo{pages}{185} (\bibinfo{year}{1989}).

\bibitem{Briggs...1995AAS...18711202B}
\bibinfo{author}{{Briggs}, D.~S.}
\newblock \emph{\bibinfo{title}{{High Fidelity Interferometric Imaging: Robust
  Weighting and NNLS Deconvolution}}}, Vol. \bibinfo{volume}{187} of
  \emph{\bibinfo{series}{American Astronomical Society Meeting Abstracts}},
  \bibinfo{pages}{112.02} (\bibinfo{year}{1995}).

\end{thebibliography}

\end{document}